\documentclass{article}
\usepackage{spconf,amsmath,graphicx}
\usepackage{soul}
\usepackage{color}
\usepackage{times}
\usepackage{epsfig}
\usepackage{epstopdf}
\usepackage{amssymb}
\usepackage{mathtools}
\usepackage{multirow}
\usepackage{enumerate}
\usepackage{enumitem}
\usepackage{makecell}
\usepackage{url}
\usepackage[noend]{algpseudocode}
\usepackage[ruled,vlined]{algorithm2e}
\usepackage{verbatim}
\usepackage[keeplastbox]{flushend}
\usepackage[utf8]{inputenc}
\usepackage[english]{babel}
\usepackage{placeins}
\usepackage{hyperref}
\usepackage{subcaption}
\usepackage{breqn}
\usepackage{hyperref}

\hypersetup{
    colorlinks=true,
    linkcolor=blue,
    filecolor=magenta,      
    urlcolor=cyan,
}

\graphicspath{ {Figures/} }

\title{DeepTalk: Vocal Style Encoding for Speaker Recognition and Speech Synthesis}

\name{Anurag Chowdhury, Arun Ross, Prabu David}
\address{Michigan State University\\\vspace{-5pt}\small{\texttt{\{chowdh51, rossarun\}@cse.msu.edu}, \texttt{pdavid@msu.edu}}}

\usepackage[pscoord]{eso-pic}
\newcommand{\placetextbox}[3]{
	\setbox0=\hbox{#3}
	\AddToShipoutPictureFG*{ \put(\LenToUnit{#1\paperwidth},\LenToUnit{#2\paperheight}){\vtop{{\null}\makebox[0pt][c]{#3}}}
	}
}
\placetextbox{.2}{0.055}{~\copyright~ 2021 IEEE. Published in ICASSP 2021}

\begin{document}
%
\maketitle


%
\begin{abstract}
Automatic speaker recognition algorithms typically characterize speech audio using short-term spectral features that encode the physiological and anatomical aspects of speech production. Such algorithms do not fully capitalize on speaker-dependent characteristics present in behavioral speech features. In this work, we propose a prosody encoding network called DeepTalk for extracting vocal style features directly from raw audio data. The DeepTalk method outperforms several state-of-the-art speaker recognition systems across multiple challenging datasets. The speaker recognition performance is further improved by combining DeepTalk with a state-of-the-art physiological speech feature-based speaker recognition system. We also integrate DeepTalk into a current state-of-the-art speech synthesizer to generate synthetic speech. A detailed analysis of the synthetic speech shows that the DeepTalk captures F0 contours essential for vocal style modeling. Furthermore, DeepTalk-based synthetic speech is shown to be almost indistinguishable from real speech in the context of speaker recognition.
\end{abstract}

\begin{keywords}
Vocal Style Modeling, Speaker Recognition, Speech Synthesis, Deep Learning
\end{keywords}

\vspace{-0.3cm}
\section{Introduction}~\label{sec:intro}
Speaker recognition is the task of determining a person's identity from their voice. The human voice as a biometric modality is a combination of physiological, anatomical, and behavioral characteristics. The physical aspects of the voice production system define the physiological characteristics of the human voice~\cite{chowdhury2020fusing}, while the prosodic (pitch, timbre) and high-level (lexicon) traits define the behavioral characteristics. A majority of automatic speaker recognition systems use only the physiological speech features due to their high discriminability and ease of characterization~\cite{voicerecoMFCC}. However, such automatic speaker recognition systems are vulnerable to audio degradations, such as background noise and channel effects~\cite{guo2017robust}. Behavioral speech characteristics, while being susceptible to intra-user variations, are considered robust to audio degradations~\cite{carey1996robust}. Behavioral features can also complement the speaker-dependent speech characteristics captured by physiological features, and can be combined to improve speaker recognition performance~\cite{adami2003modeling}. Behavioral speech features, when used judiciously, can help in the development of robust speaker recognition systems. 

A person's behavioral speech characteristics are defined by their long-term and short-term speaking habits, referred to as their `vocal style.' Long-term vocal styles are acquired over time and are influenced by social environments and native language~\cite{adami2003modeling}. Short-term vocal styles are more volatile and are influenced by the target audience (addressing a crowd versus talking over the phone) and emotional state~\cite{wu2006study}. Furthermore, apart from a speaker's idiolect, their vocal anatomy also influences their behavioral speech characteristics, thereby constraining the differences between their physiological and behavioral speech features~\cite{san2019sistema}. Behavioral speech characteristics have been used for performing speaker recognition~\cite{shriberg2005modeling}. Most of these techniques require speech data annotated at the word- and frame-level for extracting behavioral speech features~\cite{shriberg2005modeling}, posing a challenge for their development. Recently, deep learning-based methods have been developed to learn vocal style features from speech data without any word- or frame-level annotations~\cite{shen2018natural,wang2018style,jemine2019master}. However, most of these methods use handcrafted speech features, such as Mel-frequency Cepstral Coefficients (MFCC)~\cite{voicerecoMFCC}, to represent the input audio, thus contending with the vulnerabilities and performance bottlenecks of handcrafted features~\cite{chowdhury2020deepvox}.

In this work, we develop a speech encoder called DeepTalk, to capture behavioral speech characteristics directly from raw speech audio without any word- or frame-level annotations. DeepTalk's speaker recognition performance is evaluated on multiple challenging datasets. DeepTalk is further combined with classical physiological based speaker recognition methods to improve state-of-the-art speaker recognition performance in challenging audio conditions. The fidelity of DeepTalk-based vocal style features is evaluated by integrating it with a Tacotron2-based TTS synthesizer~\cite{shen2018natural} to generate synthetic speech audios. The Deeptalk-based synthetic speech audios are shown to be indistinguishable from real speech audios in the context of speaker recognition. Therefore, DeepTalk serves the dual purpose of improving both speaker recognition and speech synthesis performance.



 \begin{figure*}[t]		
	\centering
	\includegraphics[scale = 0.6]{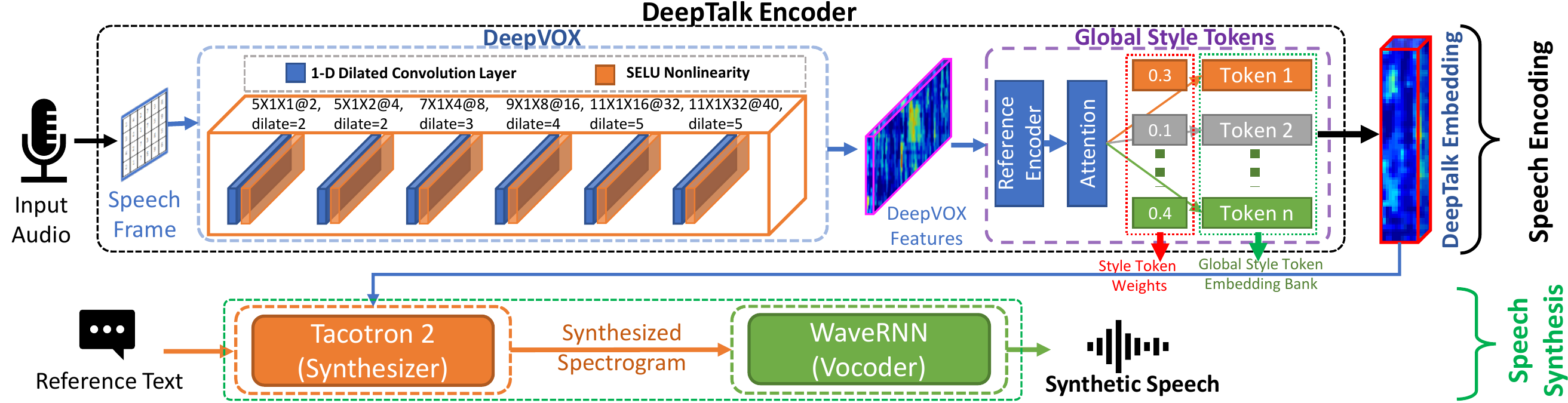}
	\vspace{-0.3cm}
	\caption{The speech encoding and speech synthesis branches of the proposed DeepTalk architecture.}
	\label{fig:arch}
	\vspace{-0.3cm}
\end{figure*}

\vspace{-0.3cm}
\section{DeepTalk}~\label{Sec:ProposedMethod}
The DeepTalk architecture (Fig.~\ref{fig:arch}) consists of separate speech encoding and speech synthesis branches.
\vspace{-0.3cm}
\subsection{Speech Encoding}
The speech encoding branch feeds a raw input audio into a DeepVOX~\cite{chowdhury2020deepvox} network to extract short-term speech features, called DeepVOX features. DeepVOX is a 1D-CNN based speech filterbank that extracts speaker-dependent speech features directly from raw speech audio. DeepVOX features are then fed to a Global Style Token (GST)-based~\cite{wang2018style} prosody embedding network to extract the DeepTalk embedding. GST uses a 2-dimentional Convolution Neural Network (2D-CNN) followed by a single-layer 128-unit unidirectional Gated Recurrent Unit (GRU) to extract a fixed dimensional reference encoding from the DeepVOX features. The reference encoding is then passed to a bank of ten randomly-initialized 128-dimensional style token embeddings called the Style Token Layer~\cite{wang2018style}. The style token embeddings serve as basis vectors of a Style Token Space representing the different vocal styles in the training data. Finally, an attention module is used to represent the reference encoding as a weighted combination of different style tokens embeddings; this is referred to as the DeepTalk embedding. We train both the DeepVOX and GST networks together using a triplet-based speaker embedding learning framework~\cite{chowdhury2020deepvox} to maximize the speaker-dependent vocal style information in the DeepTalk embedding. This allows DeepVOX to learn the speech representation best-suited for vocal style extraction using the GST network. 

\subsection{Speech Synthesis}
The speech synthesis branch feeds the DeepTalk embedding and a reference text into a Tacotron2-based synthesizer to generate a Mel spectrogram, which is then converted to the synthetic speech waveform using a WaveRNN-based neural vocoder~\cite{kalchbrenner2018efficient}. The synthetic speech's similarity to the target speaker's original speech is then qualitatively and quantitatively evaluated. The qualitative evaluation is done by manually listening to the synthetic speech and visually comparing the spectrograms of the original and synthetic speech (see Fig.~\ref{fig:spec}). The quantitative evaluation is done by comparing the two audios using speech embedding techniques (see Fig.~\ref{fig:tsne}). For training the Tacotron2 model, we use a two-phase approach. In the first phase, we train the Tacotron2 and WaveRNN models on speech audio from a large group of speakers in the VoxCeleb2~\cite{chung2018voxceleb2} dataset to learn general voice characteristics present in VoxCeleb2. In the second stage, we finetune the trained models on a small set of speech samples (30 mins. long in total) of a target speaker to enable high fidelity vocal style transfer from the target speaker to the synthetic speech. 

\begin{table*}[t]
	\fontsize{6}{8}\selectfont
	\caption{Speaker verification results using the iVector-PLDA~\textbf{[M1]}, xVector-PLDA~\textbf{[M2]}, 1D-Triplet-CNN~\textbf{[M3]}, DeepVOX~\textbf{[M4]}, DeepTalk~\textbf{[M5]}, and 1D-Triplet-CNN (DeepVOX) + DeepTalk~\textbf{[M6]} methods. Here, \textbf{P1} = VoxCeleb2, \textbf{P2} = NIST SRE 2008, \textbf{P3} = Degraded NIST SRE 2008 (Babble), and \textbf{P4} = Degraded NIST SRE 2008 (F16).}
	\centering	
    
\begin{tabular}{|c|c|c|c|c|c|}
\hline 
Exp. \# & Models & Train set/Test set & TMR@FMR=1\% & minDCF & EER (in \%)\tabularnewline
\hline 
\hline 
1 & \multirow{6}{*}{M1} & P1/P1 & 86.16 & 2.04 & 5.39\tabularnewline
\cline{1-1} \cline{3-6} \cline{4-6} \cline{5-6} \cline{6-6} 
2 &  & P2/P2 & 48.7 & 5.68 & 12.37\tabularnewline
\cline{1-1} \cline{3-6} \cline{4-6} \cline{5-6} \cline{6-6} 
3 &  & P3/P3 & 39.57 & 6.37 & 13.53\tabularnewline
\cline{1-1} \cline{3-6} \cline{4-6} \cline{5-6} \cline{6-6} 
4 &  & P4/P4 & 22.73 & 8.5 & 21.13\tabularnewline
\cline{1-1} \cline{3-6} \cline{4-6} \cline{5-6} \cline{6-6} 
5 &  & P3/P4 & 6.03 & 9.93 & 35.24\tabularnewline
\cline{1-1} \cline{3-6} \cline{4-6} \cline{5-6} \cline{6-6} 
6 &  & P4/P3 & 9.5 & 9.59 & 34.11\tabularnewline
\hline 
7 & \multirow{6}{*}{M2} & P1/P1 & 55.75 & 5.03 & 11.25\tabularnewline
\cline{1-1} \cline{3-6} \cline{4-6} \cline{5-6} \cline{6-6} 
8 &  & P2/P2 & 24.2 & 8.01 & 14.15\tabularnewline
\cline{1-1} \cline{3-6} \cline{4-6} \cline{5-6} \cline{6-6} 
9 &  & P3/P3 & 22.44 & 8.35 & 15.24\tabularnewline
\cline{1-1} \cline{3-6} \cline{4-6} \cline{5-6} \cline{6-6} 
10 &  & P4/P4 & 17.15 & 9.01 & 20.88\tabularnewline
\cline{1-1} \cline{3-6} \cline{4-6} \cline{5-6} \cline{6-6} 
11 &  & P3/P4 & 7.71 & 9.73 & 34.95\tabularnewline
\cline{1-1} \cline{3-6} \cline{4-6} \cline{5-6} \cline{6-6} 
12 &  & P4/P3 & 12.17 & 9.56 & 27.54\tabularnewline
\hline 
13 & \multirow{6}{*}{M3} & P1/P1 & 82.09 & 2.65 & 5.42\tabularnewline
\cline{1-1} \cline{3-6} \cline{4-6} \cline{5-6} \cline{6-6} 
14 &  & P2/P2 & 52.5 & 5.2 & 8.18\tabularnewline
\cline{1-1} \cline{3-6} \cline{4-6} \cline{5-6} \cline{6-6} 
15 &  & P3/P3 & 35.25 & 6.54 & 11.4\tabularnewline
\cline{1-1} \cline{3-6} \cline{4-6} \cline{5-6} \cline{6-6} 
16 &  & P4/P4 & 38.50 & 7.08 & 14.96\tabularnewline
\cline{1-1} \cline{3-6} \cline{4-6} \cline{5-6} \cline{6-6} 
17 &  & P3/P4 & 8.83 & 9.84 & 29.49\tabularnewline
\cline{1-1} \cline{3-6} \cline{4-6} \cline{5-6} \cline{6-6} 
18 &  & P4/P3 & 20.00 & 8.85 & 22.64\tabularnewline
\hline 
\end{tabular}%
\begin{tabular}{|c|c|c|c|c|c|}
\hline 
Exp. \# & Models & Train set/Test set & TMR@FMR=1\% & minDCF & EER (in \%)\tabularnewline
\hline 
\hline 
19 & \multirow{6}{*}{M4} & P1/P1 & \textbf{91.98} & \textbf{1.47} & \textbf{2.91}\tabularnewline
\cline{1-1} \cline{3-6} \cline{4-6} \cline{5-6} \cline{6-6} 
20 &  & P2/P2 & 81.05 & 2.85 & 4.45\tabularnewline
\cline{1-1} \cline{3-6} \cline{4-6} \cline{5-6} \cline{6-6} 
21 &  & P3/P3 & 70.16 & 3.51 & 7.44\tabularnewline
\cline{1-1} \cline{3-6} \cline{4-6} \cline{5-6} \cline{6-6} 
22 &  & P4/P4 & 62.4 & 4.21 & 7.25\tabularnewline
\cline{1-1} \cline{3-6} \cline{4-6} \cline{5-6} \cline{6-6} 
23 &  & P3/P4 & 15.46 & 9.25 & 22.46\tabularnewline
\cline{1-1} \cline{3-6} \cline{4-6} \cline{5-6} \cline{6-6} 
24 &  & P4/P3 & \textbf{35.05} & \textbf{7.16} & \textbf{15.22}\tabularnewline
\hline 
25 & \multirow{6}{*}{M5} & P1/P1 & 87.58 & 2.09 & 4.96\tabularnewline
\cline{1-1} \cline{3-6} \cline{4-6} \cline{5-6} \cline{6-6} 
26 &  & P2/P2 & 66.73 & 3.52 & 4.44\tabularnewline
\cline{1-1} \cline{3-6} \cline{4-6} \cline{5-6} \cline{6-6} 
27 &  & P3/P3 & 50.7 & 4.47 & 6.7\tabularnewline
\cline{1-1} \cline{3-6} \cline{4-6} \cline{5-6} \cline{6-6} 
28 &  & P4/P4 & 61.53 & 4.27 & 6.53\tabularnewline
\cline{1-1} \cline{3-6} \cline{4-6} \cline{5-6} \cline{6-6} 
29 &  & P3/P4 & 10.69 & 9.76 & 28.45\tabularnewline
\cline{1-1} \cline{3-6} \cline{4-6} \cline{5-6} \cline{6-6} 
30 &  & P4/P3 & 7.54 & 9.88 & 34.04\tabularnewline
\hline 
31 & \multirow{6}{*}{M6} & P1/P1 & 91.69 & 1.52 & 3.14\tabularnewline
\cline{1-1} \cline{3-6} \cline{4-6} \cline{5-6} \cline{6-6} 
32 &  & P2/P2 & \textbf{83.56} & \textbf{2.54} & \textbf{3.91}\tabularnewline
\cline{1-1} \cline{3-6} \cline{4-6} \cline{5-6} \cline{6-6} 
33 &  & P3/P3 & \textbf{76.86} & \textbf{3.23} & \textbf{6.14}\tabularnewline
\cline{1-1} \cline{3-6} \cline{4-6} \cline{5-6} \cline{6-6} 
34 &  & P4/P4 & \textbf{66.52} & \textbf{3.6} & \textbf{5.92}\tabularnewline
\cline{1-1} \cline{3-6} \cline{4-6} \cline{5-6} \cline{6-6} 
35 &  & P3/P4 & \textbf{17.36} & \textbf{9.15} & \textbf{21.49}\tabularnewline
\cline{1-1} \cline{3-6} \cline{4-6} \cline{5-6} \cline{6-6} 
36 &  & P4/P3 & 29.37 & 7.73 & 18.09\tabularnewline
\hline 
\end{tabular}
    \vspace{-0.3cm}
	\label{tab:spkr_veri}
\end{table*}	

\vspace{-0.3cm}
\section{Dataset and Experiments}~\label{sec:experiments}
In this section, we discuss the datasets, experimental protocols, and baseline methods used to evaluate and compare DeepTalk's speaker recognition performance. Speech data used in this work have been sampled at $8$ kHz.
\vspace{-0.3cm}
\subsection{Datasets}~\label{sec:datasets}
\vspace{-0.9cm}
\begin{itemize}[leftmargin=0cm,itemindent=.5cm,labelwidth=\itemindent,labelsep=0cm,align=left]
	\setlength\itemsep{0em}
	\item \textbf{VoxCeleb2:} We use the VoxCeleb2~\cite{chung2018voxceleb2} dataset to perform speaker recognition experiments on speech collected in unconstrained scenarios. We use speech extracted from one randomly chosen video for each of the $5,994$ celebrities in the training set and the $118$ celebrities in the test set. Speech samples longer than $5$ seconds are split into multiple non-overlapping $5$-second long speech samples.
	\item\textbf{NIST SRE 2008:} We use the NIST SRE 2008~\cite{SRE08} dataset to perform speaker recognition experiments on multilingual speech data, captured under varying ambient conditions. Additionally, we degrade samples in the NIST SRE 2008 dataset with F-16 and Babble noise from the NOISEX-92 dataset~\cite{varga1993assessment}, to increase the difficulty of the task. For our experiments, we choose speech data from the `phonecall' and `interview' speech types collected under audio conditions labeled as `10-sec', `long', and `short2' across $1336$ speakers. Speech data from a random subset of $200$ speakers is reserved to evaluate the models, while the rest is used for training. 
\end{itemize}
\vspace{-0.4cm}
\subsection{Speaker Recognition Experiments}~\label{sec:spk_ver_exp}
We perform multiple experiments (Table~\ref{tab:spkr_veri}) to evaluate and compare the speaker recognition performance of DeepTalk-based behavioral speech features with several baseline speaker recognition methods. The iVector-PLDA~\cite{dehak2011front} and the xVector~\cite{snyder2018x} algorithms are used as our first and second baseline methods, respectively, due to their robustness to channel variabilities. The MSR Identity Toolkit~\cite{sadjadi2013msr} is used to perform the iVector-PLDA~\cite{dehak2011front} experiments in this work. The PyTorch-based implementation~\cite{chowdhury2020fusing} of the xVector~\cite{snyder2018x} algorithm paired with a gPLDA-based matcher~\cite{sadjadi2013msr} is used to perform the xVector-PLDA-based experiments in this work. The 1D-Triplet-CNN~\cite{chowdhury2020fusing} algorithm is used as our third baseline, as it can extract both speech production- and speech perception-based physiological speech features. The DeepVOX~\cite{chowdhury2020deepvox} algorithm is used as our fourth baseline, as it can extract vocal source- and vocal tract-based physiological speech features. Finally, the DeepTalk and DeepVox methods are combined at a weighted score level, in a 1:3 ratio (chosen empirically), to evaluate the benefits of combining physiological and behavioral speech features.
\vspace{-0.3cm}
\subsection{Speaker Recognition Results}~\label{sec:Results}
We report the speaker verification performance (see Table~\ref{tab:spkr_veri}) using True Match Rate at a False Match Rate of $1\%$ (TMR@FMR=$1\%$), minimum Detection Cost Function (minDCF) and Equal Error Rate (EER in $\%$). The minDCF is computed at a prior probability of $0.01$ for the specified target speaker ($P_{tar}$) with a missed detection cost of $10$ ($C_{miss}$).

\begin{itemize}[leftmargin=0cm,itemindent=.5cm,labelwidth=\itemindent,labelsep=0cm,align=left]
\setlength\itemsep{-0.3em}

\item In experiments 1, 7, 13, 19, 25, and 31, the VoxCeleb2 dataset is used to perform the speaker recognition experiments on a large number of speech audios collected in unconstrained scenarios. Here, the DeepVOX method and its score level fusion with the DeepTalk method obtain comparable performance and outperform all the other methods.

\item In experiments 2, 8, 14, 20, 26, and 32, the NIST SRE 2008 dataset is used to perform the speaker recognition experiments on multi-lingual speech audios portraying challenging real-life audio conditions. Here, the score level fusion of the DeepTalk and DeepVOX methods performs the best, demonstrating its robustness to challenging audio conditions.

\item In experiments 3-6, 9-11, 15-18, 21-24, 27-30, and 33-35, speaker recognition experiments are performed on the degraded NIST SRE 2008 dataset. Here, the score level fusion of the DeepTalk and DeepVOX methods performs the best, validating its robustness to audio degradations.

\item Overall, the DeepTalk method outperforms all but the DeepVOX-based speaker recognition algorithm. This demonstrates the highly discriminative characteristics of behavioral speech features extracted by DeepTalk. The score level fusion of the DeepTalk and the DeepVOX methods further improves the speaker recognition performance across majority of the experiments. This establishes the performance benefits of combining the physiological (DeepVOX) and behavioral (DeepTalk) speech characteristics. We also performed score level fusion of the DeepTalk method with 1D-Triplet-CNN, xVector-PLDA and iVector-PLDA-based methods. Similar performance improvements were noted across a majority of the experiments, with the best results achieved by the fusion of DeepTalk with DeepVOX, followed by 1D-Triplet-CNN, iVector-PLDA, and xVector-PLDA, in that order.

\end{itemize}	
	
\begin{figure}
     \centering
     \begin{subfigure}[b]{0.23\textwidth}
         \centering
         \includegraphics[width=\textwidth]{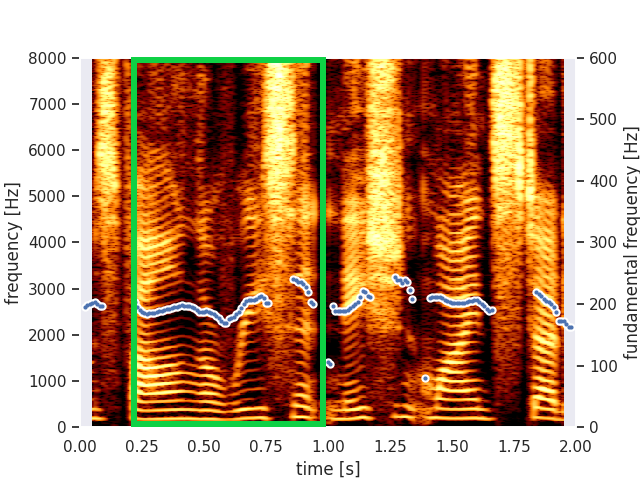}
         \caption{Original speech}
         \label{fig:spec_a}
     \end{subfigure}
     \hfill
     \begin{subfigure}[b]{0.23\textwidth}
         \centering
         \includegraphics[width=\textwidth]{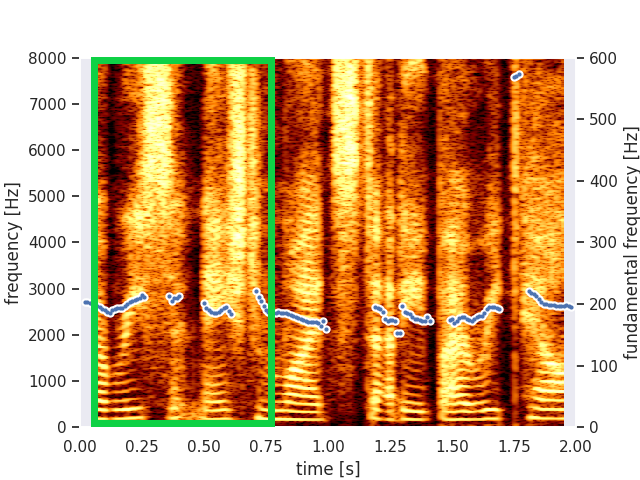}
         \caption{Baseline Tacotron2}
         \label{fig:spec_b}
     \end{subfigure}
     \hfill
     \\
     \begin{subfigure}[b]{0.23\textwidth}
         \centering
         \includegraphics[width=\textwidth]{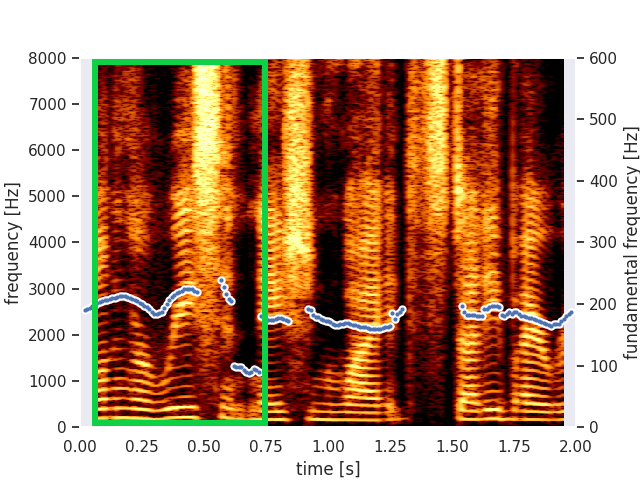}
         \caption{DeepTalk}
         \label{fig:spec_c}
     \end{subfigure}
        \caption{Spectrogram representation (overlaid with F0 contour) of a speech sample from a speaker and the corresponding synthetic speech samples generated using the baseline Tacotron2 model and the DeepTalk model. The green overlay boxes indicate the locations of corresponding speech segments across the three spectrograms. }
        \label{fig:spec}
\end{figure}

\begin{figure}[t]		
	\centering
	\includegraphics[scale = 0.8]{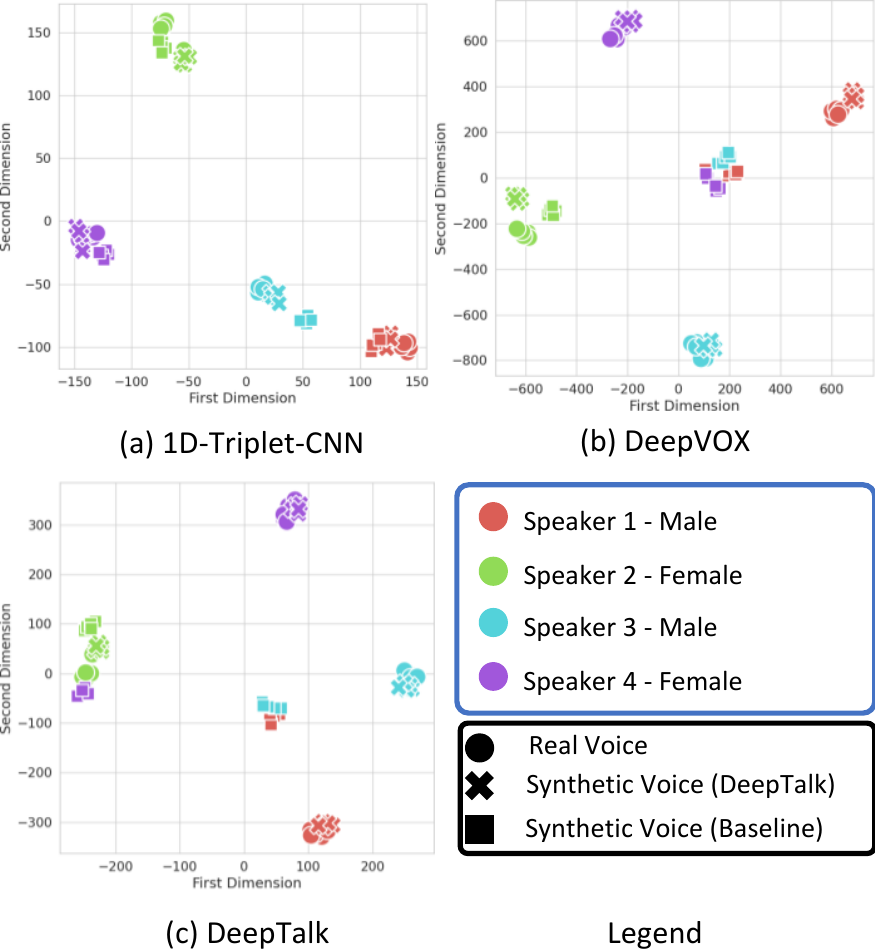}
	\vspace{-0.3cm}
	\caption{t-SNE plots of the speech embeddings of real and synthetic voice samples of four different speakers, extracted by three different speech encoders. DeepTalk's synthetic speech is embedded much closer to the real speech by all the speech encoders, as compared to the baseline synthetic speech.}
	\label{fig:tsne}
	\vspace{-0.3cm}
\end{figure}

\vspace{-0.3cm}
\subsection{Speech Synthesis Experiments and Results}~\label{sec:sp_syn_exp}
We performed multiple speech synthesis experiments, listed below, to demonstrate and analyze DeepTalk's vocal style encoding ability (Figs.~\ref{fig:spec} and~\ref{fig:tsne}). In these experiments, speech audio from the VOXCeleb2 dataset is used to train DeepTalk. The trained models were then adapted to high-quality speech audio from four different speakers (two male and two female) from the Librispeech dataset~\cite{panayotov2015librispeech} as well as internal sources. DeepTalk's speech synthesis performance was also compared to a baseline Tacotron2-based speech synthesis method~\cite{jia2018transfer}. The generated synthetic audio samples can be reviewed \href{http://iprobe.cse.msu.edu/project\_detail.php?id=20}{here}.

\begin{itemize}[leftmargin=0cm,itemindent=.5cm,labelwidth=\itemindent,labelsep=0cm,align=left]
\setlength\itemsep{-0.3em}
\item Copy synthesis experiment: Here, the DeepTalk method extracts speech characteristics from an input audio and combines it with the text transcript (of the same input sample) to recreate the input audio. The spectrogram representation of DeepTalk's synthetic speech displays greater visual similarity to the original speech, especially at frequencies higher than 2500Hz, compared to the baseline (Fig.~\ref{fig:spec}). Furthermore, the high visual similarity between the F0 contours of the original speech and DeepTalk's synthetic speech (indicated by green overlay boxes in Fig.~\ref{fig:spec}) demonstrates DeepTalk's efficacy at vocal style modeling~\cite{mary2006prosodic}.

\item Speaker Matching Experiment: Here, the 1D-Triplet-CNN, DeepVOX, and DeepTalk-based speech encoding methods extract speech embeddings from original and synthetic (both DeepTalk and baseline) speech samples for the four different speakers. The speech embeddings are then visualized using t-SNE~\cite{maaten2008visualizing}(Fig.~\ref{fig:tsne}). All the speech samples used in this experiment contain different speech utterances, ensuring a text-independent speaker matching scenario. Across all speech encoding methods, the speech samples synthesized by the DeepTalk method are embedded much closer (mean Euclidean distance of 45) to the corresponding real voice samples from the same speaker, compared to the baseline method (mean Euclidean distance of 189). This demonstrates DeepTalk's ability to generate near-indistinguishable synthetic speech samples in the context of speaker recognition. 
\end{itemize}

\vspace{-0.6cm}
\section{Ethical Implications}~\label{sec:Ethics}
In this work, we demonstrate DeepTalk's ability to reliably model the vocal style of a given speaker and transfer it to a synthetic speech with high fidelity. While this technique can improve the user-experience of Speech Generating Devices (SGD)~\cite{ross2020security} and digital voice assistants, several concerns are raised by its potential misuse for creating DeepFake speech. For example, in the past, DeepFake speech has been used to mimic an influential person's voice for defrauding~\cite{stupp2019fraudsters}. Therefore, such a technology should be used responsibly while adhering to appropriate privacy-protection laws.

\vspace{-0.3cm}
\section{Conclusion}
Behavioral speech characteristics are robust to audio degradations and complement physiological speech characteristics for voice biometrics. Therefore, it is beneficial to develop vocal style modeling techniques, such as the proposed DeepTalk algorithm, and combine it with physiological speech features for improving speaker recognition performance, as was evidenced in the experimental results. DeepTalk has also been integrated with a Tacotron2-based TTS synthesizer to generate highly realistic synthetic speech, demonstrating its efficacy at high-fidelity vocal style modeling. Therefore, it is essential to continue developing vocal style modeling algorithms and combine them with physiological speech characteristics to improve speaker verification and speech synthesis performance in challenging audio conditions.

\vspace{-0.3cm}
\section{Acknowledgement}
We thank Susi Elkins of WKAR Public Media for providing us with high-quality audio samples for finetuning the DeepTalk method. Portions of this work were funded by the National Association of Broadcasters. 

\bibliographystyle{IEEEbib}
\bibliography{strings,refs}

\begin{thebibliography}{10}

\bibitem{chowdhury2020fusing}
Anurag Chowdhury and Arun Ross,
\newblock ``Fusing {{MFCC}} and {{LPC}} features using {{1D Triplet CNN}} for
  speaker recognition in severely degraded audio signals,''
\newblock {\em IEEE TIFS}, 2020.

\bibitem{voicerecoMFCC}
Lindasalwa Muda, Begam KM, and I~Elamvazuthi,
\newblock ``Voice recognition algorithms using mel frequency cepstral
  coefficient ({MFCC}) and dynamic time warping ({DTW}) techniques,''
\newblock {\em Journal of Computing}, 2010.

\bibitem{guo2017robust}
Jinxi Guo, Ruochen Yang, Harish Arsikere, and Abeer Alwan,
\newblock ``Robust speaker identification via fusion of subglottal resonances
  and cepstral features,''
\newblock {\em The Journal of the Acoustical Society of America}, 2017.

\bibitem{carey1996robust}
Michael~J Carey, Eluned~S Parris, Harvey Lloyd-Thomas, and Stephen Bennett,
\newblock ``Robust prosodic features for speaker identification,''
\newblock in {\em ICSLP}, 1996.

\bibitem{adami2003modeling}
Andre~G. Adami, Radu Mihaescu, Douglas~A. Reynolds, and John~J. Godfrey,
\newblock ``Modeling prosodic dynamics for speaker recognition,''
\newblock in {\em ICASSP}, 2003.

\bibitem{wu2006study}
Wei Wu, Thomas~Fang Zheng, Ming-Xing Xu, and Huan-Jun Bao,
\newblock ``Study on speaker verification on emotional speech,''
\newblock in {\em ICSLP}, 2006.

\bibitem{san2019sistema}
Eugenia San~Segundo, Pedro Univaso, and Jorge Gurlekian,
\newblock ``Sistema multiparam{\'e}trico para la comparaci{\'o}n forense de
  hablantes,''
\newblock {\em Estudios de fon{\'e}tica experimental}, pp. 13--45, 2019.

\bibitem{shriberg2005modeling}
Elizabeth Shriberg, Luciana Ferrer, Sachin Kajarekar, Anand Venkataraman, and
  Andreas Stolcke,
\newblock ``Modeling prosodic feature sequences for speaker recognition,''
\newblock {\em Speech Communication}, 2005.

\bibitem{shen2018natural}
Jonathan Shen, Ruoming Pang, Ron~J Weiss, Mike Schuster, Navdeep Jaitly,
  Zongheng Yang, Zhifeng Chen, Yu~Zhang, Yuxuan Wang, Rj~Skerrv-Ryan, et~al.,
\newblock ``Natural {TTS} synthesis by conditioning wavenet on mel spectrogram
  predictions,''
\newblock in {\em ICASSP}, 2018.

\bibitem{wang2018style}
Yuxuan Wang et~al.,
\newblock ``Style tokens: Unsupervised style modeling, control and transfer in
  end-to-end speech synthesis,''
\newblock in {\em ICML}, 2018.

\bibitem{jemine2019master}
Corentin Jemine et~al.,
\newblock ``Master thesis: Automatic multispeaker voice cloning,''
\newblock {\em Universite de Liege, Liege, Belgique}, 2019.

\bibitem{chowdhury2020deepvox}
Anurag Chowdhury and Arun Ross,
\newblock ``{DeepVOX}: Discovering features from raw audio for speaker
  recognition in degraded audio signals,''
\newblock {\em arXiv preprint arXiv:2008.11668}, 2020.

\bibitem{kalchbrenner2018efficient}
Nal Kalchbrenner, Erich Elsen, Karen Simonyan, Seb Noury, Norman Casagrande,
  Edward Lockhart, Florian Stimberg, Aaron Oord, Sander Dieleman, and Koray
  Kavukcuoglu,
\newblock ``Efficient neural audio synthesis,''
\newblock in {\em ICML}, 2018.

\bibitem{chung2018voxceleb2}
Joon~Son Chung, Arsha Nagrani, and Andrew Zisserman,
\newblock ``Voxceleb2: Deep speaker recognition,''
\newblock in {\em INTERSPEECH}, 2018.

\bibitem{SRE08}
``2008 {NIST} speaker recognition evaluation training set part 2 ldc2011s07,''
  \url{https://catalog.ldc.upenn.edu/LDC2011S05},
\newblock Accessed: 2021-02-11.

\bibitem{varga1993assessment}
Andrew Varga and Herman~JM Steeneken,
\newblock ``Assessment for automatic speech recognition: {II. NOISEX-92:} a
  database and an experiment to study the effect of additive noise on speech
  recognition systems,''
\newblock {\em Speech communication}, 1993.

\bibitem{dehak2011front}
Najim Dehak, Patrick~J. Kenny, Réda Dehak, Pierre Dumouchel, and Pierre
  Ouellet,
\newblock ``Front-end factor analysis for speaker verification,''
\newblock {\em TASLP}, 2011.

\bibitem{snyder2018x}
David Snyder, Daniel Garcia-Romero, Gregory Sell, Daniel Povey, and Sanjeev
  Khudanpur,
\newblock ``{X-Vectors:} robust {DNN} embeddings for speaker recognition,''
\newblock in {\em ICASSP}, 2018.

\bibitem{sadjadi2013msr}
Seyed~Omid Sadjadi, Malcolm Slaney, and Larry Heck,
\newblock ``{MSR} identity toolbox v1.0: A {MATLAB} toolbox for
  speaker-recognition research,''
\newblock {\em Speech and Language Processing Technical Committee Newsletter},
  2013.

\bibitem{panayotov2015librispeech}
Vassil Panayotov, Guoguo Chen, Daniel Povey, and Sanjeev Khudanpur,
\newblock ``Librispeech: an {ASR} corpus based on public domain audio books,''
\newblock in {\em ICASSP}, 2015.

\bibitem{jia2018transfer}
Ye~Jia, Yu~Zhang, Ron Weiss, Quan Wang, Jonathan Shen, Fei Ren, Patrick Nguyen,
  Ruoming Pang, Ignacio~Lopez Moreno, Yonghui Wu, et~al.,
\newblock ``Transfer learning from speaker verification to multispeaker
  text-to-speech synthesis,''
\newblock in {\em NeurIPS}, 2018.

\bibitem{mary2006prosodic}
Leena Mary and B~Yegnanarayana,
\newblock ``Prosodic features for speaker verification,''
\newblock in {\em ICSLP}, 2006.

\bibitem{maaten2008visualizing}
Laurens van~der Maaten and Geoffrey Hinton,
\newblock ``Visualizing data using {t-SNE},''
\newblock {\em JMLR}, 2008.

\bibitem{ross2020security}
Arun Ross, Sudipta Banerjee, and Anurag Chowdhury,
\newblock ``Security in smart cities: A brief review of digital forensic
  schemes for biometric data,''
\newblock {\em Pattern Recognition Letters}, 2020.

\bibitem{stupp2019fraudsters}
Catherine Stupp,
\newblock ``Fraudsters used {AI} to mimic {CEO’s} voice in unusual cybercrime
  case,''
\newblock {\em The Wall Street Journal}, vol. 30, 2019.

\end{thebibliography}

\atColsBreak{\vskip5pt} 
\end{document}